\documentclass[a4paper, 11pt]{article}
\pdfoutput=1 

\usepackage{amsthm}
\usepackage{graphicx}
\usepackage{amsmath}
\usepackage{amssymb}
\usepackage{amsfonts}

\usepackage{cite}
\usepackage{color}
\usepackage{color}
\usepackage{etoolbox}
\patchcmd{\footnotemark}{\stepcounter{footnote}}{\refstepcounter{footnote}}{}{}
\usepackage[colorlinks=true,urlcolor=red,citecolor=blue,hyperfootnotes=false]{hyperref}
\newtheorem{theorem}{Theorem}[section]

\newtheorem{definition}[theorem]{Definition}
\definecolor{verde}{cmyk}{.83,.21,1,.08}
 
\newcommand{\blu}[1]{{\color{blue}\sf #1}}

\newcommand{\R}{\mathbb R}
\newcommand{\Z}{\mathbb Z}
\newcommand{\gC}{\mathbb C}
\newcommand{\be}{\begin{equation}}
\newcommand{\ee}{\end{equation}}
\def\beqa{\begin{eqnarray}}
\def\eeqa{\end{eqnarray}}
\newcommand{\eqn}[1]{(\ref{#1})}
\def\nn{\nonumber}
\newcommand\DER{{\text{\textup{Der}}}}
\newcommand\modul{{\mathbb M}}
\newcommand\caZ{{\mathcal Z}}

\def\appendix#1{\addtocounter{section}{1}\setcounter{equation}{0}
\renewcommand{\thesection}{\Alph{section}}
\section*{
\thesection\protect\indent \parbox[t]{11.715cm} {#1}}
\addcontentsline{toc}{section}{Appendix\thesection\ \ \ #1} }

\newcommand{\C}{\mathbb C} 

\def\nn{\nonumber}
\newcommand{\tr}[1]{\:{\rm tr}\,#1}
\newcommand{\Tr}[1]{\:{\rm Tr}\,#1}

\def\be{\begin{equation}}
\def\ee{\end{equation}}
\def\bea{\begin{eqnarray}}
\def\eea{\end{eqnarray}}

\newcommand{\half}{{\textstyle{1\over 2}}}

\newcommand{\del}{\partial}

\newcommand{\e}{{\mathrm e}}

\newcommand{\dd}{{\mathrm d}}

\begin{document}

\begin{titlepage}


\begin{center}

\baselineskip=24pt

{\Large\bf The Gribov problem in Noncommutative  gauge theory}

\baselineskip=14pt

\vspace{1cm}

Maxim Kurkov and Patrizia Vitale 
\\[6mm]
$^1${\it Dipartimento di Fisica ``E.~Pancini'', Universit\`{a} di Napoli
{\sl Federico II}}\\
{\it Monte S.~Angelo, Via Cintia, 80126 Napoli, Italy}
\\[4mm]
$^2${\it INFN, Sezione di Napoli}\\
{\it Monte S.~Angelo, Via Cintia, 80126 Napoli, Italy}
\\{\small\tt
 max.kurkov@gmail.com, patrizia.vitale@na.infn.it}

\end{center}

\begin{abstract}
After reviewing Gribov ambiguity of non-Abelian gauge theories, a phenomenon  related to the topology  of the bundle of  gauge connections, 
 we show that  there is a similar feature for  noncommutative QED over Moyal space, despite the structure group being Abelian,  and we exhibit an infinite number of solutions for the equation of Gribov copies.  This is a genuine effect of noncommutative geometry which disappears when the noncommutative
parameter vanishes. 
\end{abstract}

\end{titlepage}


\section{Introduction}
This article is based on a lecture given at the XXV International Fall Workshop
on Geometry and Physics in Madrid and it is aimed at illustrating the appearance of Gribov ambiguity \cite{gribov}, which is a phenomenon related to the topology of the bundle of  gauge connections,   in the framework of noncommutative gauge theory.

The Gribov ambiguity   is better understood  in the context of functional quantization of gauge theories. These are theories with first class constraints,  the generators of gauge transformations. In such a context physical degrees of freedom have to be identified with a gauge fixing procedure:  the physical  carrier space of dynamics is  defined  by  picking one representative on each gauge orbit, that is, by considering the quotient of the kinematical carrier space with respect to the gauge group.This is usually realized in the functional formalism through the Faddeev-Popov prescription. The Gribov ambiguity amounts to the fact that  there could be different field configurations which obey the same gauge-fixing condition, but which are related by a gauge transformation, that is, they  are on the same gauge orbit.    As first shown by Singer \cite{singer} and independently by Narasimhan and Ramadas \cite{narasim}, it can be given a precise mathematical characterization in the language of fiber bundles. Gribov ambiguity is a manifestation of topological obstructions to the existence of a global section for the relevant principal bundle.  

 In the first part of the  paper we   review the problem in the framework of standard gauge theory, stressing the geometric and topological issues. In the second part we  approach the problem in the framework of noncommutative gauge theory.  We first review the formulation of gauge theories in the noncommutative setting,  making use of the derivation based differential calculus. We thus analyze the equation for Gribov copies for noncommutative $U(1)$ gauge theory. The latter   is  based on the results obtained in \cite{CKRV16}.

\section{Gribov ambiguity in gauge theory} 

Let $M$ be the space-time  manifold and consider a principal fiber bundle over $M$, $P\rightarrow M$  with structure group a unitary group. $M$ is further assumed to be a  pseudo-Riemannian manifold.\footnote{ Eventually, we shall switch to positive definite metric, since functional quantization is defined  within  Euclidean quantum field theory.}  A pure  theory of fundamental interactions, without matter fields, is a theory where the dynamical fields are the gauge connections, $\omega\in\Omega^1(P)\otimes \mathfrak{g}$ with $\mathfrak{g}$ the Lie algebra of the structure group. Let $A\in \Omega^1(U)\otimes \mathfrak{g}$, $U\subset M$,  a local representative of the gauge connection and $F= dA+ A\wedge A$ the local curvature two-form. When $M$ is the Euclidean space-time the classical action describing the dynamics is 
\be
S=   \frac{1}{4}\Tr F\wedge \star_H F = \frac{1}{4}\int F^a_{\mu\nu} {F^{\mu\nu}}^a \dd^n x \label{act}
\ee
where  $F= F^a_{\mu\nu} \tau_a dx^\mu\wedge dx^\nu$, $\star_H$ is the Hodge product  and $\tau_a$ are the generators of the Lie algebra. The trace is to be intended as a scalar product over  both Lie algebra and forms. On integrating by parts we arrive at
\be
S= \frac{1}{2}\int \dd^n x\, {\int \dd^n y\,} A^a_\mu(x) M_{ab}^{\mu\nu} (x,y) A^b_\nu(y) \label{action}
\ee
with
\be 
M_{ab}^{\mu\nu} (x,y)= (-\Box \delta^{\mu\nu} +\del^\mu\del^\nu) \delta^{(n)} (x-y)\delta_{ab}. \label{Mop}
\ee
Within the functional quantization approach one defines the generating functional of Green's functions 
\be
Z[J]= \int \mathcal{D} A \e^{-\frac{1}{2}(S[A]+S_I[A,J])}  \label{zeta}
 \ee
with $S[A]$  the Euclidean action and $S_I= \tr  (J A) $. From $\ln Z[J]$ one obtains the quantum action $\Gamma[{A}]$ through Legendre transform. {The} Gaussian integral in \eqn{zeta} can be {formally} performed:
\be 
Z[J] = \left(\det M\right)^{-\frac{1}{2}} \exp\left( \frac{1}{2} \int J M^{-1} J\right) 
\ee
with $M^{-1}$ the Euclidean propagator, when  the operator $M_{ab}^{\mu\nu}$ defined by \eqref{Mop} is  invertible.
Unfortunately this is not the case for gauge theories. 

When $A$ is a $U(N)$ is the gauge connection the free action is invariant under gauge transformations
\be
A \rightarrow A^g= g A g^{-1} + dg g^{-1}  \label{gaugetransf}
\ee
with $ g:M\rightarrow U(N)$. Thus, on considering field configurations of the form  $dg g^{-1}$ (so called pure gauge terms), we have
\be
M^{\mu\nu} \del_\nu g g^{-1}= 0 \label{zeromodes}
\ee
showing that, because of gauge invariance,   the operator \eqn{Mop} has eigenvectors with zero eigenvalue (so called zero modes),  hence it is not invertible unless we perform the integral in Eq. \eqn{zeta} over equivalence classes of gauge connections.  

To this, let us define more accurately the configuration space of gauge theories and the group of gauge transformations. As above, let  $P$ a  principal $G$-bundle over  $ M$, smooth manifold representing space-time (which, rigorously, should be compact). The structure group $G$ is a finite dimensional Lie group, which we choose to be  U(N).   

\begin{definition}{\rm  An {\it automorphism} of $P$ is  a {\it diffeomorphism} $\varphi:P\rightarrow P$ which is G-equivariant, that is 
$\varphi (p\cdot g)= \varphi(p)\cdot g$ for all $p\in P$ and $g\in G$.}
\end{definition}

Every $\varphi\in {\rm Aut}(P)$ induces a diffeomorphism $\tilde\varphi$ on the basis manifold.
The map, $H$,  which associates $\tilde \varphi\in  {\rm Diff} (M)$  to $\varphi\in {\rm Aut}(P)$ is a {\it group homomorphism}.  Thus, the kernel of $H$, given by those automorphisms of $P$ which are mapped to the identity   in Diff($M$), is a group. This allows for a mathematical definition of gauge transformations:   

\begin{definition} \label{gaudef}
\rm {The gauge group of $P$ is $\mathcal{G}(P):=$ker$(H)$. Its elements are called {\it  gauge transformations} or also vertical automorphisms, because they are such that $\pi(\varphi(p))=\pi(p)$. }
\end{definition}
Gauge transformations of vector and spinor fields are implemented by the action of  $\mathcal{G}(P)$ on the vector and spinor bundles associated to $P$. 
 
An equivalent  definition, more physically oriented, is the following:  

The gauge group is homeomorphic to the group of smooth maps from space-time to the structure group $G$. 

For Euclidean space-time $\R^N$, physical considerations\footnote{See for example \cite{nair}, cap. 10  where, on coupling gauge fields to matter fields, the request of  invariance of physical states under the action of constraints,  imposes, at fixed time, that $g(x)$ tend to the identity at spatial infinity.} impose $  g(x) \rightarrow 1$ as $ |{ x}| \rightarrow \infty$ which amounts to compactify the base manifold
 \be 
 \mathcal{G}\simeq \rm{Map}(S^n \rightarrow G). \label{phydefgau}
 \ee 
The kinematical configuration space of gauge theory  is   
$ \mathcal{A}$, the  space of gauge connections of $(P,M,G)$, which are  locally represented by Lie algebra valued one-forms on the base manifold $A: M \rightarrow \Omega^1(M)\otimes \mathfrak{g}$,  transforming under the action of the gauge group according to Eq. \eqn{gaugetransf}.
Physical configurations are therefore equivalence classes with respect to the gauge transformation \eqn{gaugetransf},  which belong to the quotient  space
$
\mathcal{B}= \mathcal{A}/\mathcal{G}.
$
In order to perform the functional integral in \eqn{zeta}, one has to  integrate over $\mathcal{B}$ instead than $ \mathcal{A}$, that is, choose a representative for each equivalence class, by   fixing the gauge. 
  
Mathematically, this amounts to choose a surface $\Sigma_{f} \subset \mathcal{A}$  which possibly intersects the gauge orbits only once: a section for the principal bundle 
\be
\begin{array}{ccc}
\mathcal{A}(P) &\leftarrow& \mathcal{G} \\
\downarrow& & \\
\mathcal{B}(P) &  &
\end{array}
\ee
The choice of $\Sigma_f$ is  physically rephrased as a {gauge fixing},  for example $\del_\mu A^\mu= 0$ or, in general $f(A)= h$, for some chosen  functions $f, h$.   

Unless the bundle is globally trivial the kinematical configuration space is not a product:   $ \mathcal{A}\ne \mathcal{B} \times \mathcal{G}$, but let us assume for a moment that the equality holds. In such a case we have for the integration measures
\be
 [d \mu(\mathcal{A})]=  [d\mu(\mathcal{B})] ~ [d\mu (\mathcal{G}) ]
 \ee
 and, 
for gauge transformations close to the identity,  $U(x) \simeq {\mathbf 1}+ \alpha^a (x) \tau_a$, the integration measure over the kinematical configuration space     $ [d\mu (\mathcal{G}) ]$ can be replaced by $[d\alpha ]$.  
In order to perform a change of variables  $ [d\alpha ]\rightarrow  [d f(A) ]$, we need  the Jacobian of the transformation which is  
\be
{ \rm{Det}  \Delta_{FP} (x,y)= \rm{Det} \frac{\delta f^{a}(x)}{\delta \alpha^{b}(y)}}  \label{FPdet}
\ee
yielding
\be
[d \mu(\mathcal{A})] \rm{Det} \Delta  =  [{ [d\mu(\mathcal{B})] }  [d\alpha ]  \rm{Det} \Delta    =  { [d\mu(\mathcal{B})] }\; [d f ]
\ee
and, finally, integrating   over $[df]$ with the insertion of a delta function $\delta(f(A) -h(x))$ which implements the gauge choice, we obtain the measure on the quotient space:  
\be
  [d \mu(\mathcal{A})] \; \rm{Det} \Delta\;\; \delta(f(A) -h(x))={ [d\mu(\mathcal{B})] }. 
  \ee
  The Jacobian in \eqn{FPdet} is the so called Faddeev-Popov determinant. 
  \subsection{Gribov ambiguity}
The gauge fixing described above is not enough to remove unphysical degrees of freedom if the theory is non-Abelian. 
Indeed, let us consider the gauge orbit 
\be
A^g =g  A g^{-1} + d g g^{-1}  \simeq  A+ D\alpha
\ee
with $D\alpha=d\alpha+ \alpha{\wedge} A = d\alpha+ \alpha^a\wedge A^b [\tau_a, \tau_b]$. 
 The gauge fixing condition  $\del^\mu A^g _\mu= 0$  yields 
 \be
 \del_\mu D^\mu \alpha= 0 \label{copies}
 \ee
 which may have nontrivial solutions, whenever the gauge group is non-Abelian.\footnote{\label{footnote3} In the Abelian case we only have trivial solutions, if we further assume that $\lim_{x\rightarrow\infty} \alpha(x)=0$.} This is the so called  {\it equation of copies} and the phenomenon is known as Gribov ambiguity.  Notice that $ - (\del_\mu D^\mu) \delta^{(4)}(x-y) \delta^{ab} $ is exactly the FP determinant for this choice of gauge fixing. 

Let us return to the global approach and let us show how the existence of Gribov copies (solutions of Eq. \eqn{copies}) is the manifestation of the fact that  the bundle $\mathcal{A}\rightarrow \mathcal{B}$ is nontrivial \cite{singer,narasim}.

The kinematical configuration space
$\mathcal{A}$ is an affine space. Indeed any convex combination 
\be
 A_\tau= (1-\tau) A_1 + \tau A_2 \;\;\; 0\le \tau\le 1
 \ee
  is a gauge connection, because it satisfies
\be
A_\tau^g =g  A_\tau g^{-1} + dg g^{-1} 
\ee
therefore $\mathcal{A}$ is topologically trivial. Let us consider the gauge group
${\mathcal{G}}= \{g: S^4\rightarrow G\}$. The fundamental group 
$\Longrightarrow \Pi_1(\mathcal{G})$ may be identified with $\Pi_5(G)= \{g: S^5\rightarrow G\}$.    
\be
\Pi_1(\mathcal{G})\simeq\Pi_5(G).\label{pioneg}
\ee
 Thus we can use standard results in topology  which state that, 
for $G= U(N) $
\beqa \label{homotopy}
\Pi_5(U(N)) &=& \Z, \, N\ge 3; \nn\\
\Pi_5(U(N))  &=& \Z_2, \,N=2; \nn\\
\Pi_5(U(N))  &=& 0,\,  N=1 
\eeqa
showing that, by virtue of \eqn{pioneg}, the group manifold $\mathcal{G}$ is nontrivial except for the Abelian case. 
 Let us come to the physical configuration space 
 ${\mathcal{B}}= \mathcal{A}/\mathcal{G}$. Since $\mathcal{A}$ is homotopically trivial whereas $\mathcal{B}$ and $\mathcal{G}$ in general aren't,\footnote{On considering the long exact sequence 
$$..\rightarrow \pi_n(\mathcal{G})\rightarrow \pi_n(\mathcal{A}) \rightarrow \pi_{n} (\mathcal{B})\rightarrow \pi_{n-1(}\mathcal{G})\rightarrow... \rightarrow \pi_{0} (\mathcal{A}) $$ 
 we have that 
$\Pi_k(\mathcal{B})= \Pi_{k-1}(\mathcal{G})$}   $\mathcal{A}$ cannot be globally trivialized as the product of $\mathcal{B}$ and $\mathcal{G}$ unless $\mathcal{G}$  is topologically trivial. On the basis of \eqn{homotopy}, both $\mathcal{G}$ and $\mathcal{B}$ are only trivial for  $G=U(1)$, which is the case for electrodynamics.   

This global analysis  translates into the fact that Eq. \eqn{copies}, namely $\Box \alpha =0$,   only has trivial solutions in the Abelian  case.  
Vice-versa, we can conclude that non-Abelian gauge theories do not admit global sections, which 
 amounts to the FP operator $\Delta $  having non trivial zero modes.

\section{Noncommutative Electrodynamics  on $\R^{2n}_\theta$}
In this section we shall briefly review the formulation of  Elettrodynamics in the noncommutative setting of Moyal space-time, $\R^{2n}_\theta$.

This   is the  simplest  noncommutative space,   modeled on the phase-space of quantum mechanics,  the {\it quantum phase-space}.   
In order to define the latter, one considers the dual description of classical phase-space   in terms of its algebra of functions (classical observables) and quantizes it. The algebra of quantum observables represents quantum phase space. This is noncommutative, because the operator product is noncommutative, moreover, it has   no  underlying, dual notion of smooth manifold anymore.

Equivalently, one can describe quantum observables in terms of smooth functions on classical phase-space with a {\it noncommutative} or star product. This is the  Moyal-Weyl-Wigner description of quantum mechanics. 

Following the same approach for classical space-time, say $\R^{2n}$, one replaces $({\mathcal{F}(\R^{2n}), \cdot)}$  with a noncommutative algebra, $\R^{2n}_\theta \equiv ({\mathcal{F}(\R^{2n}), \star)}$.  The Moyal star-product is so defined:
\begin{equation}
f \star_\theta g(x) := (2\pi)^{-2n}
\int_{\R^{2n}}\int_{\R^{2n}}\,f(x + \half\theta Ju)g(x + v)
\,e^{-iu\cdot v}\,d^{2n} u\,d^{2n} v
\ee
with $J$ antisymmetric $2n\times 2n$ matrix such that $J^2=-\mathbf{1}$.  Its popular asymptotic expansion reads
\be
(f\star g)(x)=f(x)\exp \left\{ \frac{i}{2}\,\theta ^{\rho \sigma }\overset{
\leftarrow }{\partial _{\rho }}\overset{\rightarrow }{\partial _{\sigma }}
\right\} g(x) 
\ee
yielding,  for coordinate functions, 
\be x^i \star x^j - x^j\star x^i = i \theta^{ij} 
\ee
and also 
\be
x^i \star f =\blu{x^i \cdot f} +\frac{i}{2}\,\theta ^{i  j  }\del_j f 
\ee
which defines the Lie algebra of derivations $\del_j\in \DER(\R^n_\theta)$ as inner, with respect to the product: 
\be
\del_j f = i \theta^{-1}_{jk} (x^k\star f-f\star x^k) \;\;\,{\rm with }\;\; \del_j (f\star g)= \del_j f \star g+ f \star \del_j g .
\ee
The Moyal star product possesses an important property: it is cyclic and closed
namely
\be
\int d^{2n} x\, f\star g= \int d^{2n}x\, g\star f = \int d^{2n} x\,  f\cdot g \;\;
\ee
which can be shown by integration by parts.\footnote{An instance of a cyclic product which is cyclic but not closed is the Wick-Voros product. Its relation to Moyal product and its application to quantum field theory is discussed in \cite{WV}. We follow here the convention  of \cite{KV17}, so that a {\it   closed star product}  of two elements in the noncommutative algebra is a product whose integral is equal to the integral of the pointwise commutative product}
The algebraic properties of classical gauge invariant actions on  Moyal space are described by a simple version of the derivation-based differential calculus.  
The latter, introduced  long ago \cite{segal, dbv, marmolandi},  is a generalization of the de Rham differential calculus.  For mathematical details and applications to NCFT, we refer the reader to  \cite{Wallet,MVW13,GVW14}. In what follows we give a short review based on \cite{ Marmo:2004re}.
\subsection{Differential calculus for (noncommutative) associative
algebras} 
Given the commutative  associative algebra  ${\mathcal A}$ of smooth functions over a manifold $M$, the usual  differential calculus can
be equivalently defined algebraically, once a Lie algebra of
derivations, $\DER(\mathcal{A}) $, is given (see for example
\cite{segal,marmolandi}). Having defined one-forms as linear maps from 
$\DER(\mathcal{A}) $   to ${\mathcal A}$, the exterior derivative $d$ is
defined for one forms as \be d\alpha (X,Y)= X(\alpha(Y))-
Y (\alpha(X)) - \alpha([X,Y])\label{higherforms} \ee
It is easily verified that  $d^2=d\circ d$ is zero. Higher forms are
defined as skew-symmetric multilinear maps from $\DER(\mathcal{A}) $ to
the associative algebra ${\mathcal A}$. Then, the exterior derivative is easily generalized
\beqa
d\omega(X_1,..., X_{p+1})& :=& \sum_{i=1}^{p+1} (-1)^{i+1} X_i  \left(\omega( X_1,..\vee_i.., X_{p+1})\right) \\
&+&  \sum_{1\leq i < j \leq p+1} (-1)^{i+j} \omega( [X_i, X_j],..\vee_i..\vee_j.., X_{p+1}) \label{eq:koszul},
\eeqa
with $\vee_i$ meaning that the argument $i$ is omitted. 

This construction can be extended  to   noncommutative algebras,    once  we have chosen a set of derivations
of 
${\mathcal A}$, such that 
 \be X\left( f\star g\right) = \left(X
f\right) \star g + f\star\left(X  g\right), ~~~ X \in
\DER(\mathcal{A}) ,~~ f,g\in {\mathcal A} \label{std}\ee where $\star$ is the
noncommutative product in ${\mathcal A}$. 
For  Moyal algebra $\R^{2n}_\theta$  the Lie algebra of derivations is the Abelian algebra generated by the  derivatives $\partial_\mu$, $\mu=1,...,n$ . 
Zero-forms are identified with the algebra
itself , $ \Omega^0=\mathcal{A}.$  Then the exterior derivative
is implicitly defined by 
\be  df (X)=X( f)\label{d}\ee It
automatically verifies the Leibnitz rule because of Eq. \eqn{std}. Moreover $ d^2=0 $  because  $\star$-derivations close a Lie algebra. The second
step consists in defining $\Omega^1$ as a left (or right) $\mathcal{A}$-module that is \be g df(X)=g\star X( f).\ee  To
construct $\Omega^2$ we observe that  
\be df \wedge_\star  dg(X, Y) = df(X)\star df(Y) -df(Y)\star df(X) \label{west}
\ee 
where $\wedge_\star$ is the deformed wedge product. Because of noncommutativity $
df\wedge_\star dg \ne - dg\wedge_\star df. $ In a similar way to
$\Omega^1$, $\Omega^2$ is defined as a left $\mathcal{A}$-module
with respect to the $\star$-multiplication \be f dg \wedge_\star
dh(X,Y)=f\star dg(X)\star dh(Y) - f\star
dg(Y)\star dh(X).\ee Higher $\Omega^p$ are built along the same lines. 
\subsection{Gauge connection}%
 %

We then consider a natural noncommutative extension of the notion of connection, as introduced in \cite{dbv} where one replaces complex  vector bundles of physical fields over space-time, with fiber $\C^n$,   with   right-modules, $\modul$ over $\mathcal{A}$.  A connection on $\modul$ can be conveniently defined 
by a linear map ${\nabla} :  \DER(\mathcal{A})\times  \modul \rightarrow \modul$ satisfying
\begin{equation}
{\nabla}_X (m f) = mX( f) + {\nabla}_X (m) f,\ {\nabla}_{c X}( m) = c{\nabla}_X (m),\ 
{\nabla}_{X + Y} (m) = {\nabla}_X (m) + {\nabla}_Y (m) \label{connect}
\end{equation}
for any $X,Y \in \DER(\mathcal{A})$, $f \in \mathcal{A}$, $m \in \modul$, $c \in \caZ(\mathcal{A})$, the center of the algebra.  Hermitian connections  satisfy  for any real derivation $X \in \DER(\mathcal{A})$
\begin{equation}
 X(h(m_1,m_2))=h(\nabla_X(m_1),m_2)+h(m_1,\nabla_X(m_2)), \forall m_1,m_2\in\modul,\label{hermitconnect}
\end{equation}
where $h:\modul\otimes\modul\to\mathcal{A}$ denotes a Hermitian structure on $\mathcal{A}$. The curvature is the linear map $R(X, Y) : \modul \rightarrow \modul$ defined by
\begin{equation}
 R(X, Y) m = [ {\nabla}_X,{\nabla}_Y ] m - {\nabla}_{[X, Y]}m,\ \forall X, Y \in \DER(\mathcal{A})\label{courgene}.
\end{equation}

The group of gauge transformations   of $\modul$, ${\cal{U}}(\modul)$, is   defined \cite{Wallet} as the group of automorphisms of $\modul$ compatible both with the structure of right $\mathcal{A}$-module and the Hermitian structure, i.e 
\be
g(mf)=g(m)f,\;\;\; h(g(m_1),g(m_2))=h(m_1,m_2)\;\;\; \forall g\in{\cal{U}}(\modul), \;\;\; \forall m_1,m_2\in\modul \label{hermcond}
\ee
This definition is the natural algebraic counterpart of Def. \ref{gaudef}. 

For any $g\in{\cal{U}}(\modul)$ we have 
\begin{eqnarray}
{\nabla}^g_X&:&\modul\to\modul,\ {\nabla}^g_X = g^{-1}\circ {\nabla}_X \circ g\label{gaugeconnect}\\
R(X,Y)^g&:&\modul\to\modul,\ R(X,Y)^g=g^{-1}\circ R(X,Y) \circ g\label{gaugecurv}.
\end{eqnarray}

Since we shall eventually consider a gauge theory with structure  group  $U(1)$, namely  electrodynamics,   the relevant vector bundle in the commutative case  is a complex line bundle. This is generalized by means of a one-dimensional ${\mathcal{A}}$-module $\modul=\gC \otimes\mathcal{A}$. As Hermitian structure  we choose $h(f_1,f_2)=f_1^\dag f_2$ and take real derivations. Then a Hermitian connection is entirely determined \cite{Wallet} by its action on the one-dimensional basis $\nabla_X({\mathbf{1}})$. We have  $ \nabla_X(f)= \nabla_X({\mathbf{1}}) f + X(f)$,with  $\nabla_X({\mathbf{1}})^\dag=-\nabla_X({\mathbf{1}})$. This defines in turn the 1-form connection $A$  by  means of 
\begin{equation}
A:X\to A(X):=\nabla_X({\mathbf{1}}), \;\; \forall X\in\DER(\mathcal{A})
\end{equation}
From the compatibility condition with the hermitian structure, Eq. \eqn{hermcond}, one obtains  that  gauge transformations are  the group of unitary elements of the algebra. Indeed, on using $g(f)= g( \mathbf{1}  f)= g( \mathbf{1})\star f $ and imposing compatibility, we get 
$ h(g(f_1), g(f_2))= h(f_1,f_2)$ which implies $ g(\mathbf{1})^\dag\star  g(\mathbf{1}) =\mathbf{1}$.   
We pose $g(\mathbf{1})= U \in \mathcal{U} (\R^{2n}_\theta) $
  the group of unitary elements of  the algebra $\R^{2n}_\theta$, acting multiplicatively on the left of $\R^{2n}_\theta$. From   Eqs. \eqref{gaugeconnect}, \eqref{gaugecurv} we obtain
\begin{equation}
A_\mu^g=g\star A_\mu\star g^\dag+i \partial_\mu g \star g^\dag,\ F_{\mu\nu}^g=g\star F_{\mu\nu}\star g^\dag,\;\;\;\;\; \forall g\in{\cal{U}}(\mathbb{R}^n_\theta)\label{gaugetrans}
\end{equation}
where, to make contact with usual  notation, we have set  $i R_{\mu\nu}=F_{\mu\nu}=\del_\mu A_\nu -\del_\nu A_\mu -i [A_\mu,A_\nu]_\star $.  
  
  Being  unitary elements of $\mathbb{R}^{2n}_\theta$ gauge transformations may be written as star exponentials
  \begin{equation}
g[\alpha]=  \exp_{\star}\left(i \alpha\right),  \label{gtrans}
\end{equation}
and  the star exponential  is by definition 
\begin{equation}
\exp_{\star}(i\alpha) \equiv \sum_{n=0}^{\infty} \frac{(i)^n}{n!}\underbrace{%
\alpha\star...\star \alpha}_{\mbox{$n$ times}}.
\end{equation}
with $\alpha$ is some function of $x$ considered as a parameter of the
transformation. In the next section we shall study the infinitesimal form of Eq. \eqn{gtrans}.

\section{Gribov ambiguity  in noncommutative QED}

 The infinitesimal form of the gauge transformation reads 
\begin{equation}
 A^{\prime}_{\mu}[\alpha] = A_{\mu} + D_{\mu}\alpha + \mathcal{O%
}(\alpha),  \label{gtransinfin}
\end{equation}
where the appearance of the covariant derivative $D_{\mu}$, despite the gauge group  being associated to an Abelian structure group, is  an effect of non
commutativity and is given by 
\begin{equation}
D_{\mu}\alpha = \partial_{\mu}\alpha + i \left(\alpha\star A_{\mu} - A_{\mu}\star \alpha\right).
\label{cd}
\end{equation}
In the commutative limit $%
\theta\rightarrow 0$, the covariant derivative reduces to the ordinary one and the gauge transformation Eq.~\eqn{gaugetrans} gives back 
the standard Abelian gauge transformation 
\begin{equation}
A^{\prime}_{\mu}[\alpha] = A_{\mu} + \partial_{\mu}\alpha + 
\mathcal{O}(\theta)  \label{gtranscomm}
\end{equation}
\subsection{The gauge action} 
The natural generalization of the gauge  action Eq.\eqn{act} with structure group $U(1)$ to noncommutative space-time  $R^{2n}_\theta$ 
\be 
S[A]= (F,F)_\star
\ee
with a suitably defined scalar product, is obtained as follows. The wedge  product is defined for the noncommutative case  in Eq. \eqn{west}.  In turn,  the noncommutative Hodge product can be easily defined  starting from the commutative coordinate free definition 
\be
\star_H \eta= i_{\eta^\sharp} \omega, 
\ee
where $\omega$ is the volume form and $\eta^\sharp$, for each given  p-form $\eta$,  is the p-vector field associated to $\eta$ through the metric. In local coordinates,  $\eta=\eta_{j_1... j_p} dx^{j_1}\wedge ...\wedge dx^{j_p}, \; g= g^{jk}\del_j\otimes\del_k$ it reads
\be
\eta^\sharp = g[\eta]:= \eta_{j_1... j_p}  g^{rs} \Bigl(\del_r\otimes\del_s(dx^{j_1})\wedge...\wedge  \del_r\otimes\del_s(dx^{j_p})\Bigr) .
\ee 
On replacing wedge products with star-wedge products and vector fields with star-derivations we arrive at 
\be
S=\int F\wedge_\star \star_H F=  \int d^{2n} x ~~F_{\mu\nu}\star F^{\mu\nu}.
\ee
The action is easily checked to be gauge  invariant, because of  the second of Eqs. \eqn{gaugetrans} and the ciclicity of Moyal product under integration, but it yields  new pathologies with respect to the commutative case, the most studied being the Ultraviolet/Infrared mixing (UV/IR), which affects noncommutative
QFT \cite{Chep00, Min00} and, in particular, noncommutative QED \cite{Hay00,Mat00}. Such a mixing
is one of the most important open problems in noncommutative QFT as it spoils the renormalizability of the theory.   A less investigated problem is the problem of Gribov ambiguity.  Indeed it has been shown \cite{CKRV16} that noncommutative QED   similarly to commutative non-Abelian gauge theories,   exhibits Gribov copies. 

These two
fundamental problems of noncommutative QED, although at a first glance 
have nothing to do with each other, share some similarities. 
The first hint that these two apparently unrelated issues share the same
physical origin can be found in   \cite{Blaschke:2008yj}, \cite{Blaschke:2009zi}, \cite{Blaschke13}
 where the authors, adapting
an interesting result in scalar field theory \cite{Riv09}, argued that in
order to cure the UV-IR mixing in noncommutative QED one may add the term 
\begin{equation}
S_{\mathrm{fix}}\equiv \int d^{4}x\,A_{\mu }\frac{\tilde{\gamma}^{2}}{%
(-\partial ^{2})}A_{\mu }  \label{blaschke}
\end{equation}%
to the classical $U(1)$  action 
\begin{equation}
S_{\mathrm{ph}}=A_{\mu }(-\partial ^{2})A^{\mu },
\end{equation}%
that leads to the propagator of the IR-UV \emph{improved} theory 
\begin{equation}
G^{\mathrm{imp}}(p)\sim \frac{p^{2}}{p^{4}+\tilde{\gamma}^{4}}.  \label{iruv}
\end{equation}%
This has precisely the  structure of the Gribov-Zwanziger propagator introduced   to eliminate Gribov copies in the Landau gauge. 
A less investigated problem is the problem of Gribov ambiguity.  Indeed it has been shown \cite{CKRV16} that noncommutative QED   similarly to commutative non-Abelian gauge theories,   exhibits Gribov copies. For a review of noncommutative gauge theories see \cite{blaschke} and refs. therein. However, in principle, in the noncommutative case the dimensional constant $%
\tilde{\gamma}$ knows nothing about Gribov copies. Thus, unless one shows
that noncommutative geometry induces Gribov copies also in the $U(1)$ case,
the above prescription to eliminate the UV/IR mixing would appear quite 
\textit{ad hoc} and unnatural.

To this, let us choose   the Landau gauge, 
$\partial^{\mu}A_{\mu} = 0$ and replace for $A'_\mu$. 
The gauge condition $
\del^\mu A'_\mu[\alpha]= 0$ implies  the equation of copies 
\be
{\del^\mu D_\mu \alpha =0} \label{nccopies}
\ee
which  is similar  in form to the one obtained in the commutative case, Eq. \eqn{copies} and may  now have non trivial solutions, since the remark in footnote \ref{footnote3} does not apply. Let us show that, indeed,  it has an infinite number fo solutions. 

On replacing the expression of the covariant derivative and the asymptotic form of the Moyal product in Eq. \eqn{nccopies} we arrive at
\begin{equation}
-\partial^2 \alpha + \underbrace{iA_{\mu} \exp\left\{ \frac{i}{2}%
\,\theta^{\rho\sigma}\overset{\leftarrow}{\partial_{\rho}}\overset{%
\rightarrow}{\partial_{\sigma}}\right\} (\partial^{\mu}\alpha) -
i(\partial^{\mu}\alpha)\exp\left\{ \frac{i}{2}\,\theta^{\rho\sigma}\overset{%
\leftarrow}{\partial_{\rho}}\overset{\rightarrow}{\partial_{\sigma}}\right\}
A_{\mu}}_{\mbox{nonlocal terms}} = 0.  \label{zminfinfull}
\end{equation}
The presence of nonlocal terms implies that, differently form QCD, this is
not a differential equation and its resolution is a very hard task. However,
in order to say whether we have Gribov copies or not we only need to
understand whether it has nontrivial solutions $\alpha\neq0$.

After some simple manipulation and upon Fourier transformation   it is possible to recast Eq. \eqn{zminfinfull} as a  homogeneous Fredholm equation of second kind
\be
\hat{\alpha}(k)=\int d^{d}q\,\,Q(q,k)\,\,\hat{\alpha}(q) \label{zeromodeq}
\ee
with the kernel $Q$ given by 
\be
Q(q,k)=-\frac{2i\,k^{\mu }\hat{A}_{\mu }(k-q)}{k^{2}}\sin \left( \frac{1}{2}%
\,\theta ^{\rho \sigma }q_{\rho }k_{\sigma }\right) \label{iEQ}
\ee
The existence of Gribov copies has been reformulated into an eigenvalue equation for the operator $Q$.  It is possible to show that  the operator $Q$ is symmetric. In principle self adjoint operators have an infinite set of eigenfunctions
and eigenvalues, however since we are in the infinite dimensional situation
a lot depends on the properties of the kernel.  For an analysis of this equation we refer the reader to \cite{CKRV16}. Here we shall only exhibit specific gauge potentials 
  for which this equation has solutions. 
  To this, we notice that if we consider gauge potentials $\hat A_{\mu}$ which are
proportional to derivatives of $\delta(k)$ , Eq. (\ref{iEQ}) becomes a
differential equation for $\hat\alpha(k)$.

\subsection{The gauge invariant connection}\label{ginvcon}
First we try the following Ansatz 
\begin{equation}
A_{\mu} = K \theta^{-1}_{\mu\nu} x^{\nu}  \label{Acoord}
\end{equation}
with $K$ some constant to be fixed. This potential is easily verified to satisfy  the gauge fixing condition $%
\partial^{\mu}A_{\mu}$.  
In order to get rid of trivial solutions we then look for  solutions $\alpha(x)$  of \eqn{zeromodeq}
which  belong to  Schwarz space. 

The Fourier transform reads 
\begin{equation}
\hat A_{\mu}(k) = i K \theta^{-1}_{\mu\nu} \partial^{\nu}\delta(k).
\label{p1}
\end{equation}
Substituting  \eqref{p1} in the equation \eqref{iEQ} we arrive at 
\begin{equation}
-2K k^{\mu}(\theta)^{-1}_{\mu\nu} \int d^d q \,\sin{\left(\frac{1}{2}%
\theta^{\sigma\rho}q_{\rho}k_{\sigma}\right)} \hat \alpha(q)\,
^{q}\partial^{\nu} \delta(k-q) = Q k^2\hat \alpha(k),
\end{equation}
namely the following algebraic equation 
\begin{equation}
(1+ K)k^2 \hat\alpha(k) = 0,
\end{equation}
which exhibits nontrivial solutions. Indeed if and only if
$ K = -1 , $
for arbitrary even space-time dimension, any \emph{arbitrary} function $\hat\alpha(k)$ is a solution. Unfortunately,
although we found nontrivial solutions of Eq.~\eqref{iEQ}, this particular
gauge potential has a peculiar feature. One may show \cite{Wallet} that it
is invariant under gauge transformations \eqref{gtrans} and therefore we
do not have Gribov copies.

Nevertheless this potential is of interest. First of all,  the existence of such a gauge invariant connection is a purely
noncommutative feature \cite{Wallet} (also see \cite{MVW13} where such a connection has been used to study NCQED as a nonlocal matrix model) and does not exist in the commutative
limit. Second, its smooth approximations may be used in principle to search
solutions of the integral equation Eq.~\eqref{iEQ}.

\subsection{Next to the simplest situation}
 To simplify the presentation let us consider the two dimensional case.  Here we have only one noncommutative parameter, $\theta_{12}=-\theta_{21}=\theta$. The
next to the simplest gauge potential leading to a viable differential
equation is the following one:
\begin{equation}
A_{\mu}(x) \propto \theta^{-1}_{\mu\nu} x^{\nu} x^2,  \label{ntsp}
\end{equation}
which, being in two dimensions, can be further simplified to the form
\be
A_{\mu}(x)= K \varepsilon_{\mu\nu} x^{\nu} x^2,  \label{ntsp2}
\ee
with $K$ some constant to be determined and $\varepsilon_{\mu\nu}$ the Levi-Civita tensor in two dimensions.   It is easily seen to satisfy  the Landau gauge fixing
condition.  The 
corresponding Fourier transform reads 
\begin{equation}
\hat A_{\mu} (k) = i K \varepsilon_{\mu\nu}\,\square\, \partial^{\nu}
\delta(k).  \label{p2}
\end{equation}
On substituting  in the integral equation
Eq.~\eqref{iEQ} we obtain 
\begin{eqnarray}
&& K k^{\mu}\epsilon_{\mu\nu} \int d^d q \left( ^{q}\square\,
^{q}\partial^{\nu}\, \delta(q-k) \right) \,\sin{\left(\frac{1}{2}%
\theta^{\sigma\rho}q_{\rho}k_{\sigma}\right)}\, \hat \alpha(q) = \\
&&   - K k^{\mu} \varepsilon_{\mu\nu}  ^{q}\square\,
^{q}\partial^{\nu} \left[ \sin{\left(\frac{1}{2}\theta^{\sigma\rho}q_{%
\rho}k_{\sigma}\right)}\, \hat \alpha(q) \right] \bigg |_{q=k} 
= \frac{K\theta}{8}\left( \theta^2 k^4 \hat\alpha - 4 k^2 \square
\hat\alpha - 8\,\varepsilon^{\mu\nu}\varepsilon^{\eta\lambda}k_{\mu}
k_{\eta}\partial_{\nu}\partial_{\lambda}\hat\alpha\right)\nn
\end{eqnarray}
hence the zero modes $\hat\alpha(k)$have to satisfy the partial differential
equation given below: 
\begin{equation}
\left(-4k^2\square - 8\,
\varepsilon^{\mu\nu}\varepsilon^{\eta\lambda}k_{\mu}k_{\eta}\partial_{\nu}%
\partial_{\lambda} - \frac{4k^2}{Q\theta} + \theta^2 k^4\right)\hat\alpha(k)
= 0.  \label{PDE}
\end{equation}
Passing to polar coordinates  $(r,\phi)$  with 
$ 
k_1 = r \cos{\phi}\,,  
k_2 = r \sin{\phi}%
$,  Eq.~\eqref{PDE}  reads 
\begin{equation}
r^2 \hat\alpha_{rr} + 3 r\hat\alpha_r + \frac{1}{Q\theta}r^2\hat\alpha - 
\frac{\theta^2}{4}{r^4}\hat\alpha + 3\hat\alpha_{\phi\phi} = 0. \label{polcoor}
\end{equation}
which can be solved by separation of variables. 
It is shown in \cite{CKRV16} that Eq. \eqn{polcoor} admits solutions  when the amplitude $K$ takes one of the discrete values
\begin{equation}
K_{nm} = \frac{1}{\theta^2(\sqrt{3n^2 +1} + 2m+1)}, \quad n = 0, \pm1,
\pm2,...,\quad m = 0, 1, 2, ...  \label{Qnm}
\end{equation}
in such a case the general form of the zero modes is found to be \cite{CKRV16}
\be
 \hat\alpha_{nm}(r,\phi) = \left(C_1\cos{(n\phi)}
+ C_2\sin{(n\phi)} \right) r^{\sqrt{3n^2+1}-1}\exp{\left(-\frac{r^2 \theta}{4%
}\right)}\, L_m^{\sqrt{3n^2 + 1}}\left(\frac{\theta\, r^2}{2}\right)  \label{SOLUT}
  \ee
 {where} in order for $\alpha(x)$ to be real,  $ C_1,C_2,$  {are real if} $n$  {is even}
and $ C_1,C_2$ are purely imaginary if  $
n$ is odd.  $L_n^a(z)$ are the 
generalized Laguerre polynomials. 
The four-dimensional case may be  analised by a similar procedure.
 
\section{Discussion}
Having generalized the QED action to the noncommutative case of Moyal type, we have studied the equation of Gribov copies and found for simple forms of the gauge potential, an infinite number of solutions. This is a genuine noncommutative effect, which disappears when $\theta\rightarrow 0$. The role played by the matrix $\theta^{\mu\nu}$  is similar to the introduction of a background curvature of space-time, whose effect for Abelian gauge theory has been already studied in relation to  Gribov  problem \cite{decesare,canfora}. To this respect, let us notice that $\theta^{\mu\nu}$ is precisely the curvature of the gauge invariant  connection discussed in section \ref{ginvcon}, namely it behaves as a  background field affecting space-time  geometry.  It has been suggested  \cite{CKRV16} that the problem could be dealt with by a modification of the propagator, as it is done for non-Abelian gauge theories in the Gribov-Zwanziger-dell'Antonio approach \cite{GZdA}. 

The problem  shares some similarities with the UV/IR problem of noncommutative gauge theory \cite{blaschke}.
Indeed,  a propagator of the form of the Gribov-Zwanziger-dell'Antonio propagator  has
already been proposed \textit{by hand} in the NC field theory framework,
emerging from the necessity of curing the IR/UV phenomenon in scalar
translation invariant models on the Moyal plane \cite{Riv09} and it has
later been argued (see \cite{blaschke} for an up to date  review)  that the same modification could be applied to NC gauge
models, which are known to present the same kind of problem. Thus, the
Gribov-Zwanziger restriction would solve, at the same time, the problem of
the zero-modes of the noncommutative Faddeev-Popov operator and the UV/IR
mixing, clarifying the common origin of both problems.

As a final remark, it is worth emphasizing that in the scalar case where, as
we already recalled, the mixing is already present and cured through  a
modification of the action of the Gribov-Zwanziger type, there is actually a
large local symmetry of the Moyal star product at work (see \cite%
{PinzulStern} for details) which might be responsible for the existence of
copies and the demonstration could be done along the same lines as in
previous sections.
Thus, if the analysis of \cite{DZ89} can be extended to the noncommutative
case, this local symmetry of the star product could be the explanation for
the UV/IR mixing for the scalar case as well. 

\noindent{\bf Acknowledgements} 
 P.V.  acknowledges  support by COST (European Cooperation in Science  and  Technology)  in  the  framework  of  COST  Action  MP1405  QSPACE.

\end{document}